\documentclass[fleqn,twoside]{article}
\usepackage{gc}
\usepackage{amssymb}
\usepackage{amsmath}
\usepackage{amscd}
\usepackage{epsfig}

\def\Journal#1#2#3#4{{#1} {\bf #2}, #3 (#4)}

\def\etal{{\it et al.}}

\def\APO{{\em Acta Phys.Polon.} B}

\def\CQG{\em Class.Quant.Grav.}

\def\IMA{{\em Int. J. Mod. Phys.} A}

\def\JHE{\em J. High Ener. Phys.}

\def\LNP{\em Lect. Notes Phys.}

\def\JPT{\em Sov.J. Phys. JETP}

\def\NPA{{\em Nucl. Phys.} A}
\def\NPB{{\em Nucl. Phys.} B}
\def\PLB{{\em Phys. Lett.}  B}
\def\PRL{\em Phys. Rev. Lett.}
\def\PRD{{\em Phys. Rev.} D}

\def\SNP{\em Sov.J. Nucl. Phys.}

\def\al{\alpha}

\def\be{\begin{equation}}
\def\ee{\end{equation}}
\def\bea{\begin{eqnarray}}
\def\eea{\end{eqnarray}}

\def\lqcd{{\Lambda}_{QCD}}
\def\alphas{{\alpha}_s}
\def\ktwo{{\hat{\kappa}}^2}

\def\lambrs{{\Lambda}_{RS}}

\heads{Houri Ziaeepour}{QCD Color Glass Condensate Model in Warped Brane Models}

\begin{document}
\twocolumn[
\Arthead{0}{2005}{0 (0)}{0}{0}
\Title{QCD Color Glass Condensate Model in Warped Brane Models}

\Author{Houri Ziaeepour\foom 1}
{Mullard Space Science Laboratory,\\Holmbury St. Mary, Dorking, Surrey RH5 6NT, UK.}

\Abstract
{
Hadron-hadron interaction and Deep Inelastic Scattering (DIS) at very high 
energies are dominated by events at small-$x_B$ regime. Interesting and 
complex physical content of this regime is described by a phenomenological 
model called McLerran-Venugopalan Color Glass Condensate (MVCGC) model. 
The advantage of this formalism is the existence 
of a renormalization-type equation which relates directly observable 
low energy (small-$x_B$) physics to high energy scales where one 
expects the appearance of phenomena beyond Standard model. After a brief 
argument about complexity of observations and their interpretation, we 
extend CGC to warped space-times with brane boundaries and show that in a 
hadron-hadron collision or DIS all the events - and not just hard 
processes - have an extended particle distribution in the bulk. In 
other word, particles living on the visible brane escape to the bulk. 
For an observer on the brane the phenomenon should appear as time 
decoherence in the outgoing particles or missing energy, depending on the 
time particles propagate in the bulk before coming back to the brane. 
Assuming that primaries of UHECRs are nucleons, the interaction of Ultra 
High Energy Cosmic Rays (UHECRs) in the terrestrial atmosphere is the most 
energetic hadron-hadron interaction available for observation. 
Using the prediction of CGC for gluon distribution as well as classical 
propagation of relativistic particles in the bulk, we constrain the 
parameter space of warped brane models.
}
]
\email 1 {hz@mssl.ucl.ac.uk}

\section{Introduction} \label{sec:intro}
One of the wondering fact about {\it the Nature} as we know it, is the huge 
difference between the strength of the Gravity and other forces. Another 
manifestation of this puzzle is called {\it mass hierarchy} - it can induce 
large radiative corrections to the mass of Standard Model (SM) particles in 
its simplest Grand Unification (GUT) extension. In recent years the idea of 
a TeV scale gravity in the context of large extra dimensions and localized 
matter on the brane boundaries raised by N. Arkani-Hamed, \etal~\cite{nima} 
and by L. Randall and R. Sundrum~\cite{rs} inspired by some previous 
works by V. Rubakov and M. Shaposhinkov~\cite{domainwall} on domain walls in 
higher dimensional spaces and by I. Antoniadis~\cite{stringtev},P. Horava and E. Witten~\cite{ads5z2} on 
M-theory models with compactification in spaces with D-brane boundaries, have 
created lots of excitements and hope for solving this long standing puzzle.

In the first proposals it was suggested that only gravity should propagate 
in the higher dimensional bulk. Later it was proved that a total 
localization of all the fields except graviton on 3-branes is not realistic. 
In fact brane solutions are cosmologically unstable 
and at least one scalar bulk field (radion)~\cite{radion}~\cite{instab} is 
necessary to stabilize the distance between branes. In some brane models 
inflaton~\cite{infbrane} also has to propagate to the bulk to make inflation 
with necessary properties. A deeper insight to the propagation of 
gravitational waves and massive particles with bulk modes in models with 
infinite bulk has illustrated that even the warping of the bulk can not stop 
their escape from branes~\cite{geod}~\cite{shortcut}. In fact geometric 
confinement in warped models is not as efficient as it was 
expected~\cite{rs} and specially spin-1 gauge fields ~\cite {localize} can 
propagate (tunnel) to a warped infinite or macroscopic bulk ~\cite{tunnel}. 
Due to violation of the gauge symmetry~\cite{rubakovvect0}~\cite{ftrs}, 
it is practically impossible to localize gauge field completely and only 
some modification in their interactions such as enhancement of their 
coupling to localized fermions on the visible brane can partially confine 
them~\cite{escape}~\cite{escape0}~\cite{vecopac}. Even this remedy is not 
without drawbacks and the universality of fermions' charge can be 
violated~\cite{rubakovvect0}. Thus, it does not seem to be possible to 
confine all the SM fields on the brane even when the bulk metric is warped 
geometrically or just by a mild modification of the Lagrangian. The only 
way out - if branes really exist - is a symmetry which keeps at least most 
fields at energies lower than a threshold confined to the branes. This would 
be possible if branes are topological defects related to quantum gravity 
and created during a phase transition epoch in the early universe. If this 
broken symmetry prevents the production of KK-modes at low energies e.g. 
energies lower than the scale of broken symmetry, their direct production 
at present accelerator energies would be completely forbidden. For the 
same reason their effect on $Z_0/W^{\pm}$ width as well as gluon 
propagator would be extremely suppressed. For not having to add another 
scale to the theory, this scale should be close to $M_5$, the fundamental 
scale of gravity. 

Propagation of particles in the bulk has a number of cosmological 
consequences which apriori can be used to constrain brane models. 
Nevertheless, the uncertainty of cosmological measurements and the 
dependence of interpretations on the cosmological models does not yet 
permit to rule out these models completely. Thus a more controllable 
and/or close to home test is highly appreciated.

Constraints on the parameter space of brane world models from collider data 
is mainly based on the probability of direct observation of processes 
involving the production of gravitons and its Kaluza-Klein 
modes~\cite {obsgrav0} ~\cite {cosmoprob}~\cite {obsgrav2}~\cite {obsgrav3}. 
The accessible energy scales to the existent and near future accelerators is 
however limited and the fundamental scale of gravity and the effective size 
of the extra-dimensions can be constraint only up to $\sim 30 TeV$. 

Another possibility is trying to find a signature of TeV scale gravity and 
large higher dimensions in the high energy air showers produced in the 
terrestrial atmosphere by Ultra high Energy Cosmic Rays (UHECRs). Assuming 
that the primaries are single elementary particles (most probably proton or 
anti-proton), these events are the most energetic particle collisions 
available to us at present. The energy scale of these events is 
$\sim 10^{15} 
eV$ roughly 3 orders of magnitude higher than available energies even in 
near future accelerators such as LHC. If the fundamental scale of gravity 
in brane models is around Electroweak scale, UHECR 
showers with Center of Mass (CM) energies larger than $\gtrsim 1 TeV$ 
are expected to overcome symmetry restrictions and extra-dimension(s) 
should highly influence the behavior of showers.

To better understand the potential of UHRCRs as a laboratory for observing 
new physics, we should remind ourselves that there is an essential 
difference between observables in colliders and in the air-showers. In the 
former case only particles with a transverse momentum 
greater than a minimum value which depends on the detector hole size are 
detectable and the remnants of the colliding beams which include most 
energetic particles are not visible. In contrast, in an air-shower it is 
apriori possible to detect all the particles specially the most 
energetic ones and there is no discrimination between semi-hard and high 
energy remnants. Consequently, one is not restricted to see only high 
transverse momentum part of the collision. The remnant of the hadron after 
the collision consists of very energetic particles which come from the scale 
to which the hadron was boosted. They can carry important information about 
these scales which are usually unobservable in laboratory colliders.

Treating particles classically, their propagation in bulk leads to a delay 
in their arrival if they are reflected back by the gravitational warping. 
When this time delay is very short, this phenomenon is 
equivalent to a larger effective mass (heavy Kaluza-Klein modes). If however 
the time delay is long, it appears as a time decoherence in the shower - some
particles arrive detectably later than others. Due to longer fly distance of 
particles of the air showers produced by UHECRs, they are more sensitive to 
decoherence than colliders.

If during propagation into the bulk particles are considered as free, 
their wave equation can be solved and the spectrum of KK-modes can be 
determined. 
In the warped brane models the spectrum of KK-modes is very close to a 
continuum beginning from $m_5$ the 5-dim mass with a slight gap between the 
zero-mode and higher KK modes ~\cite{rs}~\cite{rubakovvect}~\cite{rubakovvect0}~\cite{escape0}~\cite{houriescape}. 
The mass of KK-modes for all types of fields has the following general 
form ~\cite {ftrs}~\cite {greenfun}~\cite {escape}~\cite{houriescape}:
\be
m_n \sim x_n \mu e^{-\mu L} \label{kkmassgen}
\ee
where $m_n$ is the mass of n$^{th}$ KK-mode, $x_n \gtrsim 3$ for $n > 0$ 
and its exact value depends on the spin of the field. Parameter $L$ is the 
effective size of the bulk. The gap between the zero-mode and higher modes 
is determined by the scale of the compactification $\mu$. Warping of the 
extra-dimension is essential for explaining the observed hierarchy between 
electroweak scale and 4-dim Planck scale. In the RS-metric, warping is the 
conformal factor $e^{-\mu y}$:
\bea
ds^2 & = & \frac {R^2}{z^2}({\eta}_{\mu\nu} dx^{\mu}dx^{\nu} - dz^2) = 
\nonumber \\
& & e^{-2\mu y}{\eta}_{\mu\nu} dx^{\mu}dx^{\nu} - dy^2 \label{rsmetric}
\eea
\be
z \equiv \frac {1}{\mu}e^{\mu y} \quad\quad {\eta}_{\mu\nu} = 
(1,-1,-1,-1)\label{zdef}
\ee
where branes are considered to be at $z = R \equiv 1/\mu$ and $z = R' \equiv 
\frac {1}{\mu}e^{\mu L}$, and $L$ is the distance between two branes. 
Warping leads to a mass spectrum for RS-type models where the KK-modes 
can be quite light:
\bea
\mu L & \sim & 36.8 - {\log}_{10}(M_5/TeV) \label{mul}\\
\mu & = & \frac {M_5^3}{M_{pl}^2}  \label{murs}
\eea
If we consider $\mu$ as a free parameter and if we don't want to add another 
high energy scale to the model, the natural choice for compactification 
scale is $M_5$ the fundamental quantum gravity scale which must be close 
to Electroweak scale $\sim 1 TeV$. It is easy to see that only if $\mu$ 
is of the order of Planck energy, the first KK-modes will not be produced 
in the present colliders ~\cite{ftrs} or interaction of UHECRs in the 
atmosphere, otherwise KK-modes are light and apriori are easily produced in 
high energy collisions.

In studying propagation of particles in the bulk we didn't consider the 
possibility and probability of production KK-modes in hadron collisions. 
Thus, question arises if the dominance of what is called small-$x_B$ events
which are soft/semi-hard, permits the production of heavy 
KK-modes~\cite{cosmoprob}. For this purpose we need a model which can 
relate the low scale observables at small $x_B$ to the high energy scales 
where new physical phenomena such as escape to the bulk can happen. Our 
argument about the need for such a model is that as we will explain in 
more details in the next section, at very high energies in majority of 
events, the signature of a new phenomenon is shrouded in low energy 
effects and we must have a suitable tool to extract the information from 
the mess of final states. 

In the following sections we first explain difficulties of analyzing the 
observations at very high energies and briefly review the QCD evolution 
models necessary for this purpose and in more details the new 
phenomenological model called McLerran-Venugopalan Color Glass Condensate 
(MVCGC) model. Then, we extend this model to warped 4+1 space-times with brane 
boundaries. And finally, we present the results for distribution of gluons in 
the bulk. To be able to have a rough estimation about the propagation time in 
more complex brane models where the application of MVCGC model is more 
difficult, we also show some results from classical propagation of 
relativistic particles in the bulk.

\section{High Energy Physics} \label{sec:hep}
Probing physics at very high energies is not an easy task. Even the 
{\it clean} signal of production of a heavy particle in the collision 
of two high energy leptons is dominated by elastic scatterings, production 
of light particles, confusions due to initial and final state soft radiative 
processes, etc. And we don't talk about technical difficulties of 
accelerating light leptons to very high energies ! It seems therefore 
that we are band to extract signals of new physics from billions of 
{\it uninteresting} interactions. 

The other possibilities - at least in the case of some models suggested 
as extension to the Standard Model - are the observation indirect effects 
and reconstruction of high energy processes by tracing back the low energy 
final particles to their high energy origin. The best example for the 
first case is the prediction of the mass of top quark from precise 
measurement of electroweak processes in LEP-1 with a center of mass 
energy much lower than the mass of top quark. The second possibility is 
more difficult and is less explored. Evidently in hadron colliders jets 
are reconstructed to obtain the rest mass of the partons which their 
hadronization make the jets and in this way it is possible to detect 
heavy particles like top quark or Higgs. 
However, one can imagine other situations. For instance, the restoration 
of symmetries broken at low energies can increase number of symmetry 
charges - approximately massless gauge bosons with roughly the same 
coupling constant as low energy bosons. This is specially expected in 
presence of Super Symmetry when all the coupling become the same. To detect 
a signature of such a process, either one should select very rare 
events which permit direct observation of production of a heavy boson. 
Or one can try to find an evidence of its existence in events where 
such bosons are produced virtually or on shell and subsequently decay to 
secondary partons which their hadronization and possible loss of a 
significant number of them out of the detector smears the signature of the 
heavy bosons. Diagrams (\ref{cleanev}) and (\ref{dirtyev}) are 
respectively examples of a {\it clean} and a {\it messy}:
\begin{figure}[t]
\begin{center}
\psfig{figure=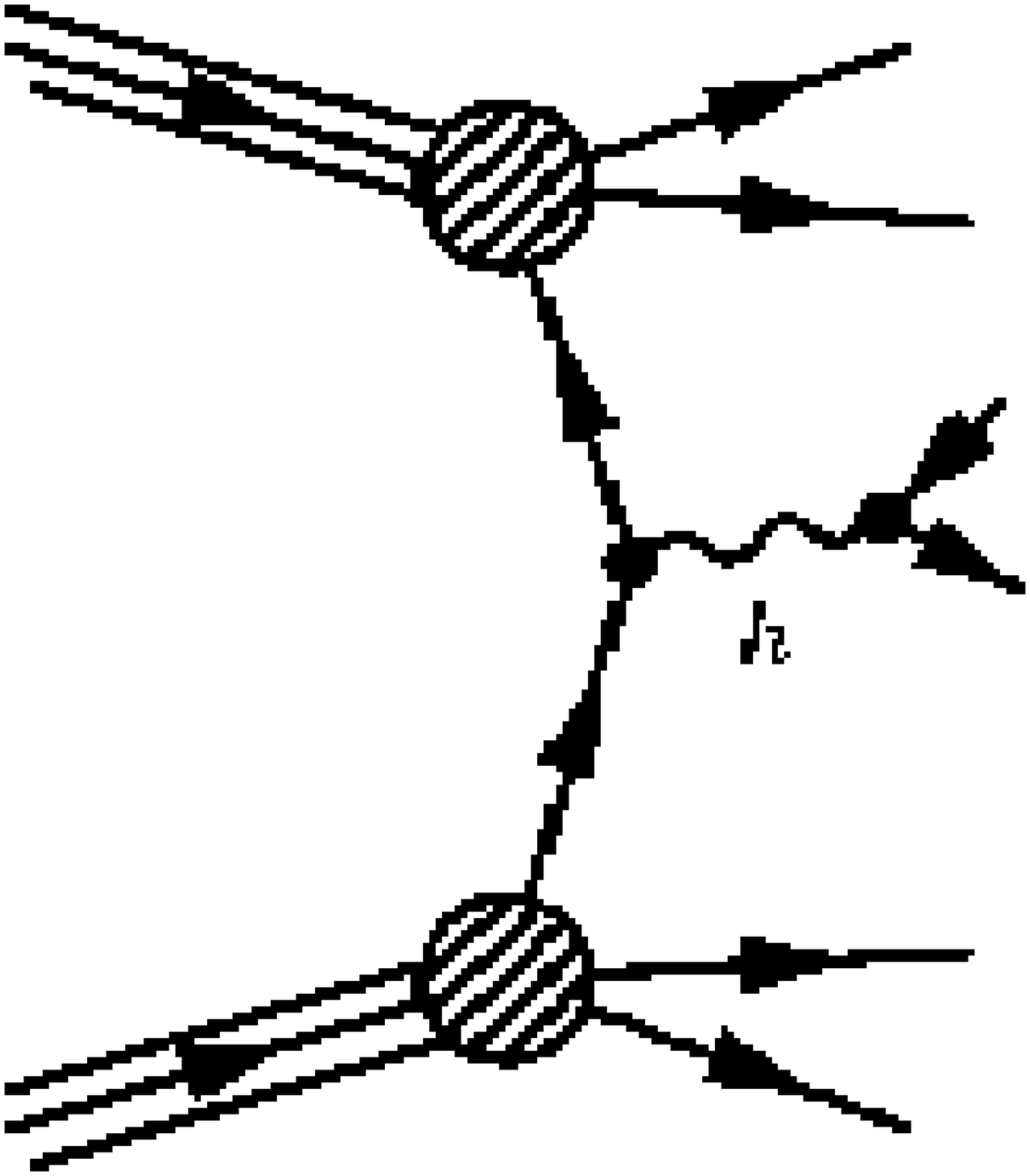,height=4cm} {(1)}
\end{center}
\end{figure} \label {cleanev}
\begin{figure}[t]
\begin{center}
\psfig{figure=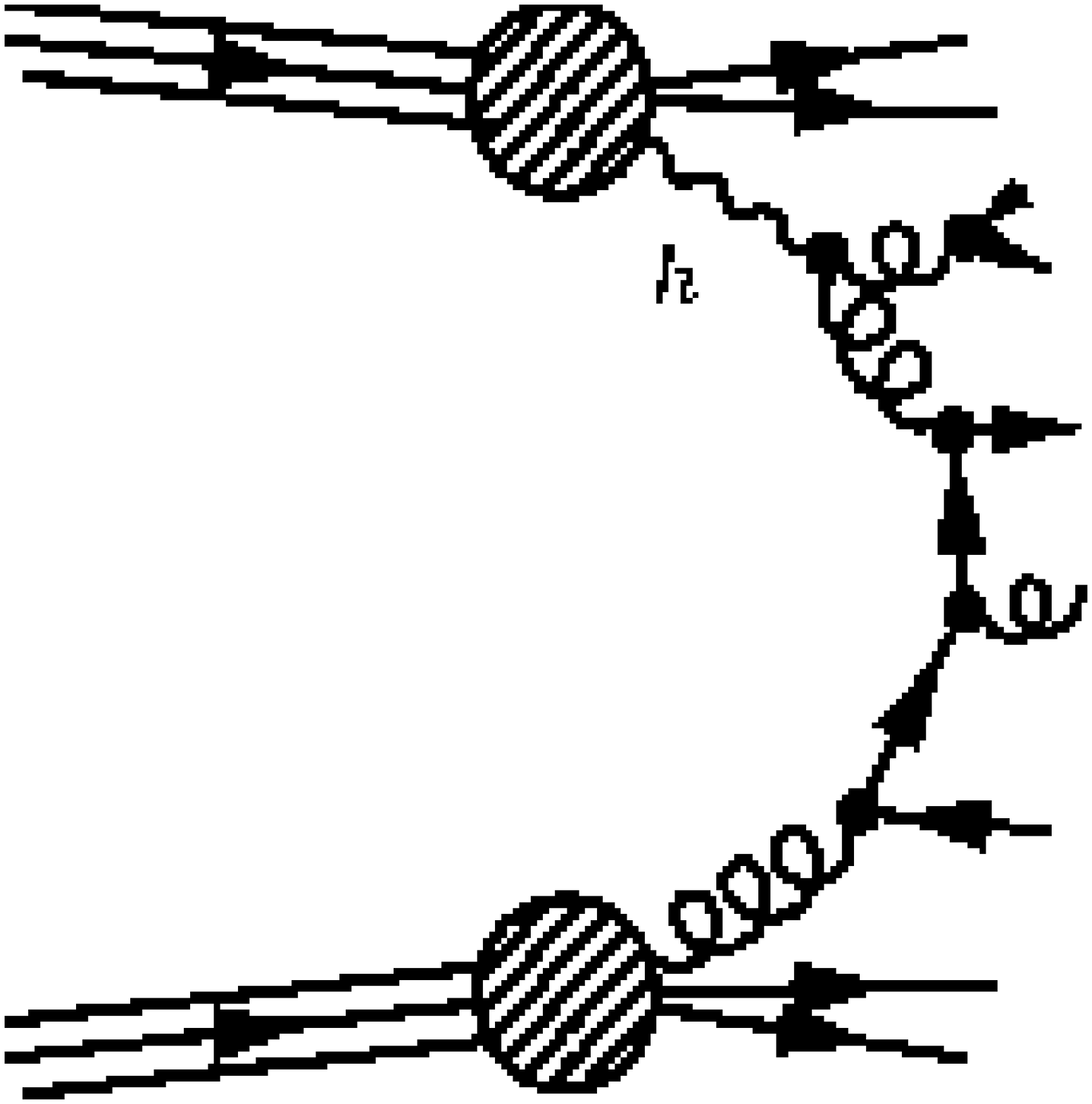,height=4cm} {(2)}
\end{center}
\end{figure} \label {dirtyev}
In the first diagram a heavy boson, labeled $h$, is produced in the 
collision of a pair of high energy collinear quark-antiquark and decays to a 
pair of quark-antiquark (or leptons). When it is close to its mass 
resonance, 
a significant enhancement in the number of events with proper rest mass is 
observed. However, the initial quarks (partons) must have enough energy to 
reach the resonance. In the second diagram the production of the heavy 
boson - here assumed to interact with gluons after restoration of a 
symmetry e.g. the broken symmetry responsible for $SU(2)-SU(3)$ splitting - 
is smeared by feather decay to gluons which in turn produce other partons. 
The boson in this diagram can be virtual and therefore can influence the 
cross-section (evidently much less than the first diagram) 
even when the available energy is not enough to produce it on-shell. 

Even in the first diagram at very high energies the quark-antiquark from 
the decay of the heavy boson can in turn have QCD radiation and make low 
transverse energy jets which also interact with the remnants of the 
colliding hadrons and make the final signature of the heavy boson very 
weak. Therefor as the borders of unknown physics approach higher and 
higher energies, the probability of {\it clean} events becomes smaller 
and observables are dominated by low energy physics. Note also that this 
depends on the nature of the extension of SM physics at very high energy 
scales and depending on the underlaying physics some effects are more 
difficult to observe than others. It is therefore necessary to find a way 
to relate low energy observables to the hidden physics behind them. 

At present, practically all the high energy colliders both man-made and 
cosmic colliders/interactions of high energy Cosmic Rays, are hadronic 
or semi-hadronic (lepton-hadron collision). It is well known 
that partons final state is always smeared by hadronization. Thus if we 
want to learn about the effect of exotic processes well before hadronization 
we must be able to relate the hadronic observables and final parton states 
to high energy partons or other particles. This is not an easy task because 
of non-perturbative interactions and multi-particle state of both initial 
and final hadrons. The common practice is finding an evolution equation 
for partons. It should permit to trace back the observed jets in the final 
state as well as the rest of the rest of the hadron bag which does not 
participate in the hard interaction, to the point where their parent 
partons have $\sim 1/3$ of hadron momentum. By approaching high energy 
scales exotic phenomena such as appearance of an extra space-time 
dimension should appear as a deviation from Standard Model. As an example 
see ~\cite{susybfkl} for evolution of gluon density in SUSY models.

In hadron-hadron and hadron-lepton Deep Inelastic Scattering (DIS), the main 
observables which determines the partial and total cross-sections are the 
density distribution of partons and its evolution at different energy scale. 
It is well known that at medium and high colliding energies these 
interactions are 
dominated by a regime called small-$x_B$ where high transverse momentum 
partons - mostly gluons - have only a very small fraction of the incident 
hadrons energy. The density of gluons in this regime is very high such that, 
although the energy scale is much higher than $\lqcd$ and $\alphas < 1$, 
${\alpha}_s \ln (1/x_B) > 1$ and perturbative expansions fail. For the same 
reason parton density evolution equations such as BFKL and DGLAP equations 
are not anymore valid approximations. The next subsection reviews very 
briefly the main aspects of these evolution models and their domain of 
validity.

\subsection{Partons Evolution Equations} \label{sec:partevol}
The non-perturbative and multi-body nature of QCD interactions in hadronic 
collisions makes the tracing of the final state particles to the parent 
partons 
very difficult. From perturbative regime of the QCD, we know that structure 
functions - presenting the distribution of partons at each energy scale - 
have an approximate scaling properties. Their main variation is with respect 
to Bj\"orken parameter $x_B$ which can be interpreted as the fraction of 
initial hadron momentum carried by a parton. The QCD radiation corrections 
add a slow energy variation proportional to $\ln Q^2$ (at lowest order) to 
the structure functions. Parameter $Q^2$ is the absolute value of the 
square of invariant energy-momentum of the exchange boson. More generally, 
it can be any energy describing the scale in which QCD interactions 
are happening. When very high energy colliders were not yet available, the 
dependence on the transverse momentum $k_{\bot}$ was usually ignored. In 
fact in DGLAP approximation (see below) structure functions are integrated 
over $k_{\bot}$:
\bea
f_a (x_B, Q^2) & = \int_{k^2_{\bot} < Q^2} dk^2_{\bot} {\phi}_a (x_B, 
k^2_{\bot}, \ln Q^2) \label{kinteg} \\
& a = \mbox{gluon, quarks, antiquark.} \nonumber
\eea
The reason for this mostly historical definition has been that at low 
energies and lowest order of perturbative QCD, 
$k^2_{\bot} \gtrsim {\Lambda}^2_{QCD}$ and 
without QCD radiation correction, its maximum value is limited by $Q^2$. 
At high energies however, $Q^2$ and $k^2_{\bot}$ become unrelated and the 
latter is controlled by the initial energy of the incident hadron.

The core part of the structure functions at low energies is non-perturbative 
and must be determined experimentally. Although due to non-perturbative 
effects the evolution with respect to $\ln Q^2$ and $k^2_{\bot}$ can not 
be calculated from first principals of the QCD, good 
approximations can be found by using factorization of perturbative 
diagrams from non-perturbative regime and some statistical arguments. 

The most popular evolution equations (also by historical order) are DGLAP 
and BFKL. As we mentioned in the previous paragraph, DGLAP equation 
describes the evolution of structure functions with $\ln Q^2$:
\bea
& &\hspace{-0.8cm}\frac {d}{d\ln Q^2} f_a (x_B, Q^2) = -f_a (x_B, Q^2) \nonumber \\
& &\hspace{-0.5cm}\sum_{b} \int_0^1 dx~x \widetilde {W}_{a \rightarrow b}(x) + 
\sum_{b} \int_{x_B}^1 \frac {dy}{y} \widetilde {W}_{b \rightarrow a}(x/y) 
f_b (y, Q^2) \nonumber \\ 
& & \label{dglap}
\eea
The kernel $\widetilde {W}_{a \rightarrow b}$ is the probability for parton 
$a$ to change to parton $b$ by a perturbative QCD interaction at a given 
$x_B$ and a given energy scale $Q^2$. The first and second terms in 
(\ref{dglap}) present respectively the probability of annihilation and 
creation of parton $a$ in perturbative QCD interactions. 

At high energies $Q^2$ and $k^2_{\bot}$ are decoupled and the latter becomes 
a better indicator of QCD interaction scale. 
BFKL equation determines the evolution of unintegrated structure function 
${\phi}_g (x_B, k^2_{\bot})$ of gluons which are the dominant parton with 
respect to $\ln (1/x_B)$. Although BFKL equation is more suitable for 
determining the density of gluons at very small $x_B$ regime of high energy 
colliders, the calculation and interpretation of its kernel function is more 
subtle and the inclusion of higher QCD orders into its kernel is more 
difficult. 

Since HERA data at very small $x_B$ became available, it has been realized 
that both DGLAP~\cite{dglap} and BFKL equations are inadequate for explaining 
the observed 
total cross-section and its evolution with kinematic parameters. Moreover, 
both in HERA and in RHIC at very small $x_B$, the cross-section which 
according to BFKL should exponentially increase, approaches a saturation and 
becomes much smaller than expected~\cite{smallxrev}. 

The main reason for the failing of DGLAP and BFKL is that none of them deal 
correctly with momentum ordering~\cite{ccfm}. In DGLAP the integration over transverse 
momentum is equivalent to considering only collinear part and momentum 
ordering is completely washed out. For large $x_B$ the effect is negligible 
as the collinear component is much larger than transverse momentum. But at 
small $x_B$ the ordering both in emitted and transferred partons becomes 
important. BFKL add the transverse momentum to the calculation of the 
kernels but assumes that all of them are of the same order (a regime 
called {\it multi-Regge})~\cite{ccfm}.

Two strategies have been followed to find structure functions and evolution 
equations enough precise, specially at small $x_B$, to explain observations. 
The first one called CCFM evolution equation~\cite{ccfm} is in the same 
spirit as 
DGLAP and BFKL models and uses the factorization of soft partons but takes 
into account the ordering of real and virtual momentums and leads to an 
evolution equation which at large $x_B$ is correct to collinear 
approximation (like DGLAP) and at small $x_B$ is not restricted to this 
approximation. Nonetheless, it has to make some simplifying 
assumptions about matrix elements in the kernel without which it is not 
possible to find a simple recurrent relation and thereby a simple evolution 
equation. Parton distributions are unintegrated similar to (\ref{kinteg}) 
but also depend explicitly on $Q^2$. The kernel (splitting function) 
includes form factors for IR regularization of soft on-shell and virtual 
gluons, latter is particularly important at low $x_B$. The differential 
representation of this evolution equation is the 
following~\cite{ccfm}~\cite{smallxrev}:
\bea
& &\hspace{-1cm}\frac{\bar{q}^2 \partial}{\partial\bar{q}^2}\frac{{\mathcal A}(x_B,
\vec{k}_{\bot},\bar{q};\vec{k}_{0\bot},Q,\mu)}{{\Delta}_s (\frac{\bar{q}}{z},
Q)} = \nonumber \\
& &\hspace{-0.5cm}\int_{x_B}^1 \frac{dz}{z} \frac{\widetilde{P}(z,\frac{\bar{q}}{z},k_{\bot})}{{\Delta}_s (\frac{\bar{q}}{z},Q)}{\mathcal A}(\frac{x_B}{z},
\vec{k}'_{\bot},\frac{\bar{q}}{z};\vec{k}_{0\bot},Q,\mu) \label{ccfm}
\eea
where $\vec{k}'_{\bot} = \vec{k}_{\bot} + \frac{1-z}{z}\vec{q}$. The 
splitting function of CCFM evolution equation $\widetilde{P}$ is complex 
and depends on the kinematic of emitted partons:
\bea 
& &\hspace{-1cm}\widetilde{P} (z,q,k_{\bot}) = \prod_i \widetilde{P}_i(z_i,q_i,k_{i\bot}) \label {splitfunc} \\
& &\hspace{-1cm}\widetilde{P}_i(z_i,q_i,k_{i\bot}) = \nonumber \\
& &\hspace{-0.5cm}\frac{\bar{\alpha}_S (k^2_{i\bot})}
{z_i}{\Delta}_{ns} (z_i,q_i,k_{i\bot}) + \frac{\bar{\alpha}_S ((1 - z_i)^2 
q_i^2)}{1-z_i} \label{splitfunci} \\
& &\hspace{-1cm}{\Delta}_s (\bar{q},Q) = \nonumber \\
& &\hspace{-0.5cm}\exp \biggl (-\int_{Q^2}^{\bar{q}^2} 
\frac{dq^2}{q^2} \int_0^{1 - Q / q} dz \frac{\bar{\alpha}_S(q^2 (1-z)^2)}
{1-z} \biggr ) \label{deltasodakov} \\
& &\hspace{-1cm}{\Delta}_{ns}(z_i,q_i,k_{i\bot}) = \nonumber \\
& &\hspace{-0.5cm}\exp \biggl (-\bar{\alpha}_S 
\int_{z_i}^1 \frac{dz'}{z'}\int_{Q^2}^{\bar{q}^2} \frac{dq^2}{q^2}
\Theta (k_{i\bot} - q)\Theta (q - z'q_i) \biggr )\nonumber \\
& & \label{deltanonsodakov}
\eea

The second method is quite different in spirit and is based on a 
phenomenological modeling of many-particle QCD processes. The initial idea 
raised by L. McLerran and R. Venugopalan~\cite {mvtheory} in the context of 
determination of
partons density in a large nuclei and is based on a classical treatment of 
color fields. Later however it was observed that it can be also applied to 
other situations such as small $x_B$ regime in collision of high energy 
partons where the density of gluons is very large and although in 
perturbative regime of QCD, large probability of interaction make the 
non-perturbative effects important. For this reason this model is now called 
{\it Color Glass Condensate (CGC)}. The reason for words {\it Color} and 
{\it Condensate} in this naming is clear. The word {\it Glass} is used 
because gluons at small $x_B$ are produced from gluons with high $x_B$ i.e. 
ones with a larger share of the energy and momentum of the initial hadron. 
In infinite momentum frame they have a time dilatation which will be 
transferred to lower energy scale partons. They evolve slowly 
compared to natural time scale, similar to a glass behavior~\cite{mvqobs}.
This observation is very important for extending this model to higher 
dimension space-times which should manifest only at very high energy scales. 

It is important to mention that the condensation of gluons in this model is 
only the asymptotic state of gluon matter at low energy scales and high 
densities, and the model can be well used at higher energy scales where the 
density of gluons is not enough high to make a condensate. This can be 
seen from the possibility of finding BFKL approximation at the lowest order 
in this model.

\begin{figure}[t]
\begin{center}
\psfig{figure=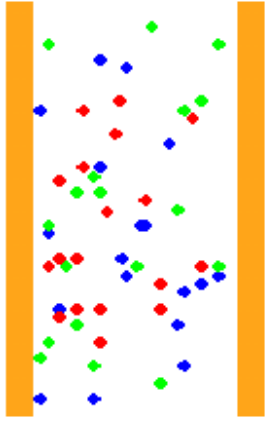,height=6cm}
\caption {Collision of two hadrons. The color/grey level of the dots is an illustration 
of color charge of the gluons. The thickness of sheets presents the scale 
${\Lambda}^+$ for which the color charge is defined. \label{fig:twodarons}}
\end{center}
\end{figure}

The idea of this model comes from the observation that a hadron with large 
momentum in the rest frame of the observer is contracted in the direction of 
the boost and its content is concentrated on a sheet-like surface. 
If $E \gg \lqcd$, partons are freed from confinement and can move in the 
direction of the boost. Due to interaction with the color sheet and 
themselves, their lifetime is however very short with respect 
to the time variation of the fast partons on the sheet. When two such 
hadrons with opposite momentums collide with each other, just before the 
collision most of the swarm of slow partons (mostly gluons) is concentrated 
between two sheets made by fast color charges of each hadron 
(see Fig.\ref{fig:twodarons}).
In a classical field theoretical  view, one can assume the sheets as an external 
charge configuration playing the r\^ole of a source for the QCD gauge field 
(gluon field) propagating between two hadron sheets. This is a QCD analog of a 
capacitor in 
electromagnetism. One can therefore solve the field equation and determine 
the field strength (gluon density) for this configuration. The main 
difference between this model and its electromagnetic analog is that in 
the latter case the charge density of the sheets is well defined. In QCD 
the charge depends on the energy scale, in other word depends on how close 
or far the observer is looking at one of the sheets (hadrons) because 
relating a charge to the sheet or to the swarm is not unique; closer an 
observer looks at the hadron, more of gluons s/he is seeing belong to the 
swarm in the space between sheets and less to the charge sheets~\cite{mvqobs}.

This phenomenological illustration is very helpful when we want to extend 
hadron-hadron collision to the very high energy scales where exotic processes 
such as extra-dimensions can be accessible to partons~\cite{houribraneqcd}. 
For instance, if hadrons 
are boosted such that the physical width of the hadron sheets becomes 
comparable to the effective size of the bulk, at very high energy scales 
when slow partons leave the charge sheet, they can enter to the bulk - 
assuming that at these scales there is no difference (symmetry breaking) 
which makes the visible brane the prefered direction. According to 
{\it Glass} concept explained before, the partons at high energies are the 
parents of low energy, small $x_B$ partons and thus by escaping to the bulk 
naturally they take with them the low energy partons. This leads to a 
significant reduction of number of partons on the brane which must be 
observable. Even when gravitational force of the warping can bring back 
the particles to the brane, a time decoherence - equivalent to a small mass 
excess should be 
observable. Notice that this picture is Lorentz invariant. To an observer 
in the frame of one of the hadrons, partons of the other hadron come from 
very short distances and can participate in exotic processes including 
emission to extra-dimension.

Although it is possible to add the effect of extra-dimensions with non-flat 
metrics to parton evolution equation such as BFKL and CCFM, we have 
prefered to use the CGC model for this purpose. The reason is the 
consistent structure of this model which, as we will show in the next 
section, is manifestly Lorentz invariant. Moreover, at least formally, 
there is no need for many simplification assumptions in the intermediate 
steps which is the case for all the evolution equation formalisms. 

\section{Color Glass Condensate (CGC) in 4+1 Space-time with Brane 
Boundaries} \label {sec:cgc}
In this section we use light-cone coordinates defined as:
\be
x^+ = x^0 + x^3 \quad , \quad x^- = x^0 - x^3 \label{lccoord}
\ee
In McLerran-Venugopalan approximation, QCD interactions are modeled by an 
effective $SU (3)$ gauge field $A^B$ of gluons. For simplicity we neglect 
the color index except when its explicit indication is necessary. In Light 
Cone (LC) gauge $A^+ = 0$. The time variation scale of 
color charge density on the sheet produced by the incoming hadron is much longer 
than the gluon swarm time scale. Therefore, it can be considered as a static 
charge. The classical dynamic equation of the effective gluon field is:
\bea
[D_A,F^{AB}](x) & = &\delta^{B+}{\mathcal W} (x^+, \vec {x})
{\rho}_{\Lambda}^+(\vec {x}) {\mathcal W}^{\dagger} (x^+, \vec {x}) \label{dyneqgen} \\
{\mathcal W} (x^+, \vec {x}) & = & T \exp \biggl \{ ig \int_{x^+_0}^{x^+} 
d{\eta}^+ \frac {R^2}{z^2} A^-({\eta}^+, \vec {x})\biggr \} \nonumber \\
& & \label{wline} \\
D_A & \equiv & {\partial}_{;A} - igA^a_A T^a \label{covder}
\eea
The symbol $;$ means covariant derivation with respect to RS metric. Color 
charge $\rho$ is $+$ component of color current:
\be
J^A (x) = {\delta}^{A+} \rho (x^-, x^{\bot}, z) \label{colorcurr}
\ee
The Wilson-line term in the right hand side of (\ref{dyneqgen}) guarantees 
the gauge invariance of this equation. The time ordering operator $T$ 
operates in (\ref{covder}) on $A^- \equiv A^-_aT^a$ and orders fields 
from right to left in increasing sequence of $x^+$. 

The subscript 
${\Lambda}^+$ of the color charge density means that it is defined at 
scale ${\Lambda}^+ = x_B P^+$ where $P^+$ is the energy of the initial hadron.
The scale ${\Lambda}^+$ indicates the maximum LC momentum of the parton swarm. 
In classical MV model it is not possible to relate models at different 
scales. But when quantum corrections are added, one can determine the 
evolution of parton density with scale ${\Lambda}^+$ and thereby relate the 
gluon distribution at $x_B \rightarrow 0$ to $x_B \rightarrow 1$ and vice 
versa. Physical quantities such as 2-point correlation functions at 
classical (tree) level are obtained from solutions of (\ref{dyneqgen}) 
integrated over all possible $\rho (x)$ with probability distribution 
$W_{{\Lambda}^+} [\rho]$. 

We assume a universal coupling in the bulk and on the branes. The aim of 
having additional coupling on the branes in ~\cite{escape} is the request 
for confinement of vector fields on the brane when fermions are confined. It is 
however easy to see that in this model fermions can't be confined to the 
visible brane because this violates the gauge invariance and makes the 
model inconsistent.

In LC gauge equation (\ref{dyneqgen}) has a solution with $A^- = 0$ in 
addition to the LC gauge condition $A^+ = 0$~\cite{jmsmallx}. The solution for other 
components is:
\be
\mathcal{A}_i = \frac {i}{g}{\mathcal U}(\vec {x}){\partial}_{;i}
{\mathcal U}^{\dagger}(\vec {x}) \quad\quad i = x^{\bot}, z  \label{asollc}
\ee
with ${\mathcal U}$ an element of QCD $SU(3)$ group. Unfortunately in this gauge 
there is no analytical way to relate ${\mathcal U}$ to the charge $\rho$ and 
to find out its explicit form. But in covariant gauge one can find an explicit 
solution. If we perform a gauge rotation to covariant gauge, $A^B$ becomes:
\be
\tilde {A}^B = {\mathcal U}^{\dagger} A^B {\mathcal U} + \frac {i}{g}
{\mathcal U}^{\dagger}{\partial}^B_;{\mathcal U} \label{covgauge}
\ee
In LC gauge $A^+ = 0$ and $\tilde {A}^+$ has K\"ahler potential form:
\be
\tilde {A}^+ = \frac {i}{g}{\mathcal U}^{\dagger}{\partial}^+_;
{\mathcal U} \label{cova}
\ee
and equation (\ref{dyneqgen}) reduces to:
\bea
-{\partial}_{;i}{\partial}_;^i \tilde {A^+} & = & \tilde {\rho} (\vec {x}) 
\quad\quad i = x^{\bot}, z \label{dyneqcov} \\
\tilde {\rho} (\vec {x}) & \equiv & {\mathcal U}^{\dagger}(\vec {x}) {\rho} 
(\vec {x}) {\mathcal U} (\vec {x}) \label{covrho}
\eea
$\tilde {\rho}$ is color charge density in covariant gauge. Using 
(\ref{dyneqcov}), equation (\ref{cova}) can be inverted and:
\be
{\mathcal U} (x^-, x^{\bot}, z) = P \exp \biggl \{ ig \int_{x^-_0}^{x^-} 
d{\eta}^- \frac {R^2}{z^2} \tilde {A}^+({\eta}^-, x^{\bot}, z)\biggr \} 
\label{rhocov}
\ee
where $P$ orders $\tilde {A}^+ \equiv \tilde {A}^+_a T^a$ from right to 
left in increasing or decreasing order of $x^-$ argument respectively for 
$x^- > x^-_0$ or $x^- < x^-_0$. We need both LC and covariant gauge 
formulations because in the latter gauge the solution of (\ref {dyneqgen}) is 
simpler, but the gluon distribution function has a simpler description in LC 
gauge.
 
Solution of (\ref{dyneqcov}) gives the propagator (Green function) in the 
warped 4+1 space-time. After defining the Fourier transform of $\tilde {A}^+$:
\be
\tilde {A}^+ (k^{\bot}, z) = \int d^2x^{\bot} e^{i k_{\bot}x^{\bot}} 
\tilde {A}^+ \label{afourier}
\ee
one can expands equation (\ref{dyneqcov}) and obtain the equation for the 
propagator:
\be
z^2 {\partial}^2_{z}\hat\Delta (z,z') + z {\partial}_{z}\hat\Delta (z,z') + 
(k^2 z^2 - 1)\hat\Delta (z,z') = R^2 \delta (z - z') 
\label{greenbessel}
\ee
with $k^2 = -k_{\bot}^2$. General solution of (\ref{greenbessel}) is:
\be
\hat\Delta (z,z') = C(z') J_1 (kz) + D(z') N_1 (kz)   \label {besselsol}
\ee
where $J_1$ and $N_1$ are respectively Bessel function of first and second 
type. Note that the momentum of partons in the direction of the initial 
hadron i.e. ${\Lambda}^+ = x_BP^+$ appears only in the scale of the model 
or equivalently in the color charge $\rho$. Also the dependence on $x^-$ is not 
dynamical which means that it should be fixed when boundary conditions are 
imposed. Integration constants $C(z')$ and $D(z')$ are determined by applying 
boundary conditions on the branes at $z = R$ and $z = R'$. This results to 
following mass spectrum for KK-modes for large $n$:
\be
m_n \approx R'^{-1}(\frac {3 \pi}{4} \pm n\pi) \label {kkmass}
\ee
Therefore the coefficient $x_n$ in (\ref{kkmassgen}) for lightest modes is 
$\sim 2.36$. The mass of all the gluon KK-modes in this model 
is real. The spectrum begins from a massless zero-mode and there is a gap 
between zero-mode and higher modes proportional to ${\mu}' \equiv R'^{-1}$ 
which for all macroscopic bulk is very small. In fact for large 
$\mu L \gtrsim 30$, even for compactification scale $\mu \sim M_{pl}$, the 
mass of lightest KK-modes is much smaller than CM energy of UHECR 
interaction in the atmosphere and when the interaction scale 
${\Lambda}^+ \gtrsim m_n$, KK-modes can be produced. 

There is also a zero-mode with $k_{\bot} = 0$ with $\hat\Delta (z,z')$ 
satisfying:
\be
z{\partial}_z (z {\partial}_z \hat\Delta (z,z')) - 1 = 
\frac {R^2}{z^2}\delta (z - z') \label{zeromodeeq}
\ee
\be
\hat\Delta (z,z') = C_0(z')z + D_0(z')z^{-1}   \label {zeromodesol}
\ee
After applying the boundary conditions to (\ref{zeromodesol}), one finds 
that the integration constants $C_0(z')$ and $D_0(z')$ are in general 
non-trivial and therefore the zero-mode propagates to the bulk but its 
wave function exponentially decreases inside the bulk. This is a well 
known fact~\cite{localize} and contrasts with scalar field case in which 
the Green function of the zero-mode is $\propto \delta (z - z')$ and 
consequently induces a discontinuity in the spectrum and separates 
zero-mode from higher KK-modes.

Propagation of gluon swarm to the bulk in MVCGC model has important 
implication for attempts to localize massless gauge bosons by adding an 
induced kinetic term on the branes. It has been shown 
~\cite{rubakovvect}~\cite{ftrs} that this term increases the coupling of the 
zero-mode to itself and if fermions are confined to the brane, the gauge 
vector field becomes quasi localized. Here it is easy to see that due to $z$ 
dependence of $\tilde {A}^+$ and therefore ${\mathcal U} (\vec {x})$, 
according to (\ref{covrho}), $\tilde {\rho}$ will depend on $z$ even when 
$\rho$ is confined to the brane. We will see later that because of the 
renormalization group equation which relates $W_{{\Lambda}^+} [\rho]$ the 
distribution 
of color charge density at each ${\Lambda}^+$ scale to other scales, $\rho$ 
or $\tilde {\rho}$ gradually receives a $z$ dependence, i.e. the color 
charge escape to the bulk even if initially - at large ${\Lambda}^+$ - it is 
concentrated on the brane. A more physical image of this process is obtained 
if one considers the transverse momentum ordering. At smaller $x_B$'s 
partons have larger transverse momentum because they have lost their energy 
by QCD radiation which at the same time increases their transverse momentum. 
When an extra space dimension is available, with each radiation the 
transverse momentum of the outing parton in the extra-dimension direction 
increases. 

Finally gluon distribution can be calculated from the matrix elements 
$\langle {\mathcal A}_a^i {\mathcal A}_b^j\rangle$ after quantizing the 
classical solution ${\mathcal A}_a^i$. The definition of the gluon 
distribution function (structure function) in LC gauge ~\cite{gdistmuller} 
can be extended to 4+1 space-times and results to the following expression:
\bea
& &\hspace{-0.7cm}x_B G(x_B, Q^2, z) = 2J_1(1) P^+ x_B Q^2 \int d^3\vec{x} 
e^{iP^+ x_B x^-} x^{\bot 2} \nonumber \\
& &\hspace{-0.5cm}\int d^3\vec{x'} \int_0^L dy'e^{\mu y'} \biggl
\langle{\mathcal A}^i (x^+, \vec {x'}, y'){\mathcal A}_i (x^+, \vec {x} - 
\vec {x'}, y - y')\biggr\rangle  \nonumber \\
& & \label {gdistfin}
\eea
In this definition of gluon distribution function $Q^2$ is the 3-dim 
transverse momentum scale and $x_B G (x_B, Q^2, z)$ is the density of
soft/semi-soft gluons up to $|k_{\bot}| < Q$. Due to the bounded non-flat 
geometry of the bulk we can not extend $|k_{\bot}|$ to the whole transverse 
directions. The brackets $\langle\rangle$ in (\ref{gdistfin}) present both 
quantum mechanical expectation function and averaging over all possible 
configuration of color charge density $\rho (x)$. The probability 
functional $W_{{\Lambda}^+} [\rho (x)]$ for each configuration at a given 
scale ${\Lambda}^+$ can be obtain from path integral quantization of MVCGC 
model. It has been proved that $W_{{\Lambda}^+} [\rho (x)]$ evolution 
satisfies a renormalization group equation~\cite{mvq}:
\bea
\frac {\delta W_{\tau}[\rho]}{\delta \tau} & = & {\alpha}_s \biggl \{ 
\frac {1}{2} \frac {{\delta}^2}{\delta {\rho}^a_\tau (x^{\bot}, z) 
\delta {\rho}^b_\tau (x^{'\bot}, z')} [W_{\tau}{\chi}^{ab}] - \nonumber \\
& & \frac {\delta}{\delta {\rho}^a_\tau (x^{\bot}, z)} [W_{\tau}{\sigma}^a]
\biggr \} \label {rge}
\eea 
where $\tau = \ln (P^+/{\Lambda}^+)$. Indexes $a$ and $b$ are color 
indexes. Matrix elements ${\sigma}^a$ and ${\chi}^{ab}$ are defined as:
\bea
{\sigma}^a & = & \langle \delta {\rho}^a \rangle \label {sigadef} \\
{\chi}^{ab} & = & \langle \delta {\rho}^a \delta {\rho}^b \rangle 
\label {chiabdef}
\eea
where $\delta {\rho}^a$ is the fluctuation of $\rho$ distribution from its 
classical value. Lowest order QCD diagrams which contribute to charge 
variation are discussed in ~\cite{mvq}.

In Gaussian approximation (see (\ref{chidef}) below) the standard deviation 
${\sigma}^2_{{\Lambda}^+} \sim T_{ab}{\chi}^{ab}$ and at lowest order (BFKL 
approximation)~\cite{mvq}:
\bea
{\sigma}^a & \approx & -g^2 D^+_y Tr \biggl (T^a G^i_{0i} (x, y) \biggr ) 
\biggl |_{x = y} \label {sigbfkl} \\
{\chi}_{ab} & \approx & 4ig^2 {\mathcal F}^{+i}_{ac}(x) 
\langle x|G_{0ij}|y \rangle {\mathcal F}^{+j}_{cb}(y) \label {chibfkl} \\
{G_0^{ij}}^{-1} (x) & = & g^{ij} {\partial}_{;B} {\partial}_;^B \quad , 
\quad i = \bot , z  
\eea
where $G_0^{ij}$ is the propagator of free gluons. In equation 
(\ref{chibfkl}) ${\mathcal F}^{+i}_{ac}$ is the gluon classical field strength 
of solution of (\ref{dyneqgen}). This approximate evaluation of $\chi$
can be used in (\ref{appaa}) below to determine the gluon distribution up to 
BFKL approximation. It can be shown~\cite{houribraneqcd} that in this 
approximation the matrix elements $\langle {\mathcal A}_a^i {\mathcal A}_b^j\rangle$ have the following expression:
\bea
& & \hspace{-1cm}\biggl\langle {\mathcal A}_a^i (\vec{x}, z) {\mathcal A}_{bi}
(\vec {x'}, z')\biggr\rangle = \nonumber \\
& & \hspace{-0.3cm}{\delta}_{ab} \chi (\vec {x}, \vec {x'}, 
z, z'){\partial}_;^i{\partial'}_{;i} \gamma (x^{\bot}, x'^{\bot}, z, z') + 
\ldots \label {appaa}
\eea
where:
\bea
\chi & \equiv & \frac {(z z')^{\frac{1}{2}}}{R} \int_{-\infty}^{max(x^-,
x'^-)} dx''^- {\sigma}_{{\Lambda}^+}(x''^-, x^{\bot}, z) \nonumber \\
& & {\sigma}_{{\Lambda}^+}(x''^-, x'^{\bot}, z') \label{chidef}
\eea
and:
\bea
& &\hspace{-1cm}({\partial}_{;k}{\partial}_;^{'k})^{-2} {\delta}^3 (x^{\bot}-x'^{\bot}, 
z, z') \equiv \gamma (x^{\bot}, x'^{\bot}, z, z') = \nonumber \\
& &\frac {1}{(2\pi)^2} \int_{{\Lambda}_{QCD}} d^2k{\bot} e^{-ik_{\bot} 
(x^{\bot}-x'^{\bot})} \nonumber \\
& & \int_R^{R'} \frac {dz''}{R^4}\Delta (z'', 
z,k^{\bot})\Delta (z'', z', k^{\bot}) + \gamma (0) \label {gammadef}
\eea
The standard deviation of the Gaussian distribution ${\sigma}_{{\Lambda}^+}$ 
can be replaced by BFKL approximation (\ref{sigadef}).

\begin{figure}[t]
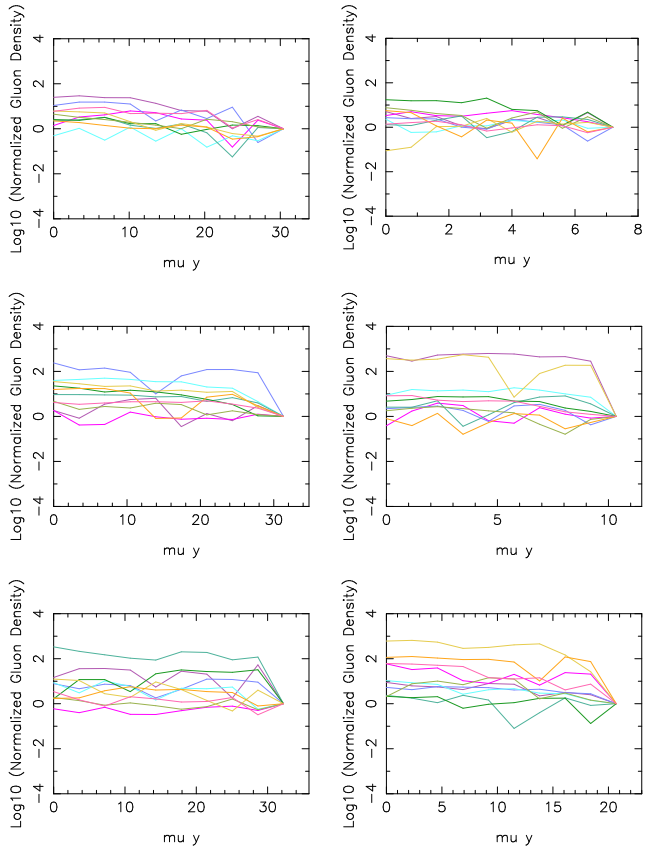

\begin{center}
\begin {tabular}{cc}
\psfig{figure=gl-11ll33.8+11nn1.5rsmunorm.eps,height=4cm,angle=-90} &
\psfig{figure=gl-11ll8+11nn1.5rsmunormmod.eps,height=4cm,angle=-90} \\
\psfig{figure=gl-14ll34.8+14nn1.5rsmunorm.eps,height=4cm,angle=-90} &
\psfig{figure=gl-14ll11.5+14nn1.5rsmunormmod.eps,height=4cm,angle=-90} \\
\psfig{figure=gl-17ll35.8+17nn1.5rsmunorm.eps,height=4cm,angle=-90} &
\psfig{figure=gl-17ll23+17nn1.5rsmunormmod.eps,height=4cm,angle=-90}
\end {tabular}
\caption{Gluon distribution in the bulk normalized to the amplitude of the 
distribution on the visible brane at $z = R'$ for Randall-Sundrum 
models for $2 \times 10^8 eV \leqslant |k_{\bot}| \leqslant 1.26 \times 
10^{10} eV$. Curves are rainbow color/grey coded. From bottom to top 
$M_5 = 10^{13} eV, 10^{14} eV$ and $10^{15} eV$, with $\mu$ obtained from 
fine-tuned RS model (\ref{murs}) i.e. $\mu = 10^{-17} eV, 10^{-14} eV$ and 
$10^{-11} eV$ respectively. For left plots 
$\log (R'/R) = \log (M_{pl}/ M_5)$, i.e. the same as (\ref{mul}). 
In right panel a smaller 
$R'/R$ is used to see the effect of a smaller bulk with same fundamental 
gravitation and compactification scale. Our tests show that apparent 
vibration of the distribution is mainly due to low resolution of our 
numerical calculation.
\label {fig:rs}}
\end{center}
\end{figure}

The result of numerical calculation of the gluon distribution for the fine-tuned 
RS model and when the bulk scale $\mu$ is of the same order as fundamental 
gravity scale $M_5$ are respectively shown in Fig.\ref{fig:rs} and 
Fig.{fig:rshighmu}. They shows that in both cases the probability of gluon 
emission into the bulk is much larger than their emission into the brane. 
Approximations we made to obtain these distributions have a validity domain 
roughly the same as BFKL approximation. Nonetheless, they can be evolved to 
lower scales by using (\ref{rge}). Even without any additional calculation 
and just by remembering that gluons at lower scales are produced from 
gluons at higher scales, one can conclude that most of the softer gluons 
should also be emitted in the bulk. Thus, if $M_5 \lesssim 10^{15} eV$, one 
had to observe a much lower flux for the high energy wing of UHECRs. In fact 
not only it does not seem to be the case, the observed flux is much higher 
than classical expectations and either a top-down model of a decaying heavy 
dark matter or extremely exotic astronomical sources are needed to explain 
their flux.

\begin{figure}[t]
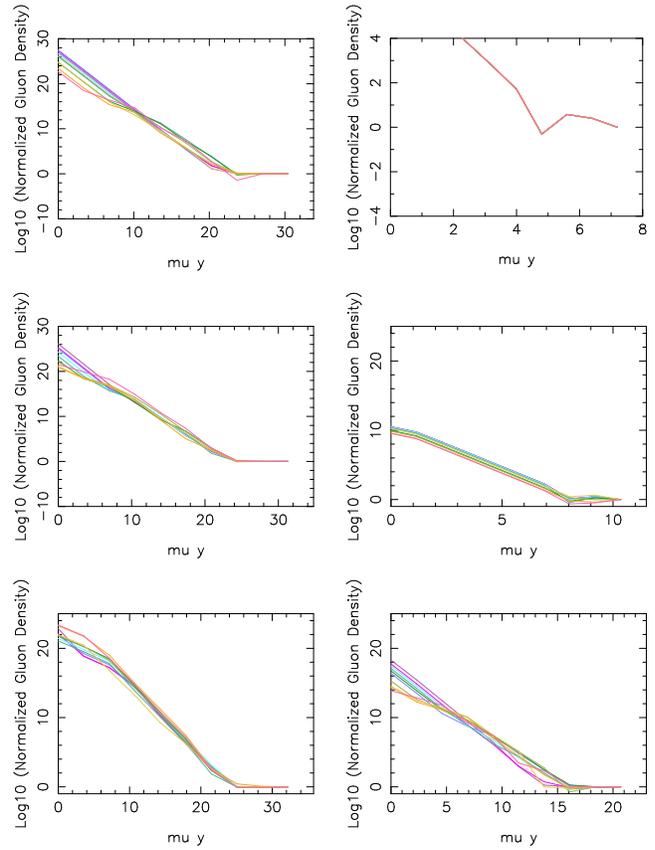

\begin{center}
\begin {tabular}{cc}
\psfig{figure=gl15ll33.8-15nn1.5norm.eps,height=4cm,angle=-90} &
\psfig{figure=gl15ll8-15nn1.5norm.eps,height=4cm,angle=-90} \\
\psfig{figure=gl14ll34.8-14nn1.5norm.eps,height=4cm,angle=-90} &
\psfig{figure=gl14ll11.5-14nn1.5norm.eps,height=4cm,angle=-90} \\
\psfig{figure=gl13ll35.8-13nn1.5norm.eps,height=4cm,angle=-90} &
\psfig{figure=gl13ll23.-13nn1.5norm.eps,height=4cm,angle=-90}
\end {tabular}
\caption{The same as Fig.\ref{fig:rs} but with $\mu = M_5$. Other details 
are the same as Fig.\ref{fig:rs}. In the top right plot the difference 
between curves for different $k^{\bot}$ is much smaller than the resolution 
of this figure and they are overlapped. \label {fig:rshighmu}}
\end{center}
\end{figure}

\section{Classical Propagation of Relativistic Particles in the Bulk}
Now that we have shown that high energy particles in a hadron-hadron 
collision penetrate to the bulk, we would like to use the classical 
propagation of these particles in the bulk to constrain other brane model. 
The reason is that although the MVCGC formalism or other evolution models 
such as CCFM can be extended to any brane model, for more complex metrics 
the calculation becomes very difficult. For a simple order of magnitude 
estimation of parameter space, a time of flight estimation can be adequate. 
The general form of the metric of a 4+1 dimensional space-time with curved 
bulk is:
\be
ds^2 = n^2 (t, y) dt^2 - a^2 (t, y) \delta_{ij} dx^i dx^j - b^2 (t, y) dy^2. 
\label {nmetric}
\ee
In a gauge where $b (t, y) = 1$, after solving Einstein equations and 
applying boundary conditions on the branes, one obtains the following 
solutions for $a (t, y)$ and $n (t, y)$:
\be
a^2 (t, y) = {\mathcal A}(t)\cosh (\mu y) + 
{\mathcal B}(t)\sinh (\mu y) + {\mathcal C}(t) \label {aa}
\ee
\bea
\dot {a}^2 (t, y) & = & n^2 (t, y) a_0^2 (t) = \nonumber \\
& & \frac {\biggl (\dot{\mathcal A}(t)\cosh (\mu y) + 
\dot{\mathcal B}(t)\sinh (\mu y) + \dot{\mathcal C}(t)\biggr )^2}
{4 a^2 (t, y)} \nonumber \\
& & \label {dotaa} \\
{\mathcal A}(t) & = & {a_0}^2 (t) - {\mathcal C(t)} \label {aaa}\\
{\mathcal B}(t) & = & -{\rho}'_{b_0} {a_0}^2 (t) \label {bbb}\\
{\mathcal C}(t) & = & -\frac {2 {{\dot {a}}_0}^2 (t)}{{\mu}^2} \label {ccc}\\
\mu & \equiv & \sqrt {\frac {2 \ktwo}{3} |{\rho}_B|} \\
n (t, y) & = &\frac {\dot {a} (t,y)}{\dot {a}_0 (t)} \label {ndef}
\eea
For any density $\rho$, ${\rho}' \equiv \rho / \lambrs$, 
$\lambrs \equiv 3 \mu / \ktwo$. The densities ${\rho}'_{b_0}$ and ${\rho}_B$
are respectively effective total energy density of the brane at $y=0$ and 
the bulk. We consider only AdS bulk models with 
${\rho}_B < 0$. The constant $\ktwo = 8 \pi / M_5^3$ is the gravitational 
coupling in the 5-dim. space-time. The model dependent details such as how 
${\rho}'_{b_0}$ and ${\rho}_B$ are related to the field contents in the 
bulk and on the brane and how they evolve are irrelevant for us as long as 
we assume a quasi-static model. The solution (\ref{aa}) is valid both for one 
brane and multi-brane models. The only difference between them is in the 
application of Israel junction conditions~\cite{kantres}~\cite{houribr}. After changing the variable $y$ to $z = e^{\mu y}$ and using (\ref{ndef}):
\bea
\dot {a}^2 (t, z) & = & -\frac {{\mu}^2 {\mathcal C}(t)}{2z} {\mathcal D}
(t, z) \label {addef}\\
{\mathcal D} & \equiv & \frac {1}{2} \biggl [(1-{\rho}'_{b_0} - 
\frac {{\mathcal C}(t)}{a^2_0}) z^2 + \frac {2 {\mathcal C}(t)}{a^2_0}z + 
\nonumber \\
& & (1 + {\rho}'_{b_0} - \frac {{\mathcal C}(t)}{a^2_0}) \biggr ] 
\label {ddef}\\
\frac {dz}{dt} & = & \mu \sqrt {{\mathcal D} (z - \frac {\epsilon}
{{\theta}^2}{\mathcal D})} \label {eqz}
\eea
$\theta$ is an integration constant. If an ejected particle to the bulk comes 
back to the brane, its velocity in the 
bulk must approach to zero at its turning point before the particle arrives 
to the bulk horizon (if it is present). The roots of (\ref{eqz}) correspond 
to these turning points and determine the propagation time in the bulk. 
The typical propagation time we are interested in is much shorter than the 
age of the Universe and therefore ${\mathcal A}, {\mathcal B}, {\mathcal C}$ 
and $\dot {\mathcal A}, \dot {\mathcal B}, \dot {\mathcal C}$ during 
propagation are roughly constant. The right hand side of (\ref{eqz}) depends 
only on $z$ and is easily integrable:
\be
{\Delta} t_{propag} \equiv 2 (t_{stop} - t_0) = \int_{z_0}^{z_{stop}} 
\frac {2 dz}{\mu \sqrt {{\mathcal D} (z - \frac {\epsilon}
{{\theta}^2}{\mathcal D})}} \label {tstop}
\ee
In (\ref{tstop}), $t_0$ is the initial time of propagation in the 
extra-dimension and $t_{stop}$ is the time when the particle's velocity 
changes its direction, i.e. when $dz/dt = 0$. The integral in (\ref{tstop}) 
is related to the elliptical integrals of the first type 
${\mathcal F} (\omega, \nu)$ where $\omega$ and $\nu$ are analytical 
functions of the denominator roots in (\ref{tstop}) and 
$z_0$. Note that $z_{stop}$ corresponds to the closest root to $z_0$.

Using this simple model, we determine the propagation time of relativistic 
particles in the bulk for some of the most popular brane models. This 
time delay must be smaller than the time resolution of air-shower detectors 
as no time decoherence - delay between arrival of most energetic particles - 
has been observed. 

\begin{figure}[t]
\begin{center}
\psfig{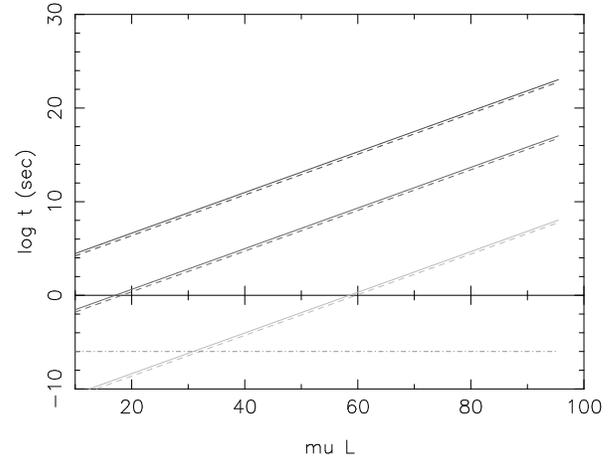}
\caption{Propagation time for relativistic particles with Lorentz factor 
$u^0_L (t_0)/N = 10^3$ (full line) and $u^0_L (t_0)/N = 1.2$ (dash 
line) in RS model. From top to buttom curves correspond to 
$M_5 = 10^{13} eV$, $M_5 = 10^{15} eV$ and $M_5 = 10^{18} eV$ (or $\mu \sim 
10^{-17} eV$, $\mu \sim 10^{-11} eV$ and $\mu \sim 10^{-2} eV$ for 
fine-tuned model) respectively. The horizontal 
line shows the time coherence precision of present Air Shower detectors.
\label {fig:rstime}}
\end{center}
\end{figure}

Fig.\ref{fig:rstime} shows ${\Delta} t_{propag}$ as a function of $\mu L$ 
and $M_5$ for massive relativistic particles. With present air shower 
detectors time resolution of order $10^{-6} sec$, only when $M_5 \gtrsim 
10^{18} eV$, the model is compatible with the observed time coherence of the 
UHE showers. For fine-tuned RS model $\mu \approx G /\ktwo$ ~\cite{rs} i.e. 
${\mu} = G M_5^3 \sim 10^{-3} eV$ for $M_5 \sim 10^{18} eV$ ~\cite{rs}. 
Due to smallness of $\mu$ and consequently lightness of KK-modes for SM 
particles even this model with large $M_5$ has already been ruled 
out~\cite {ftrs} unless a conserved quantum number prevents the production 
of KK-modes~\cite{universal}. For massless particles:
\be
z (t) - z_0 = \mu (t - t_0)^2 \label {zrs}
\ee
In (\ref{zrs}), $z (t)$ is monotonically increasing and there is no stopping 
point. With our approximations there is no horizon in the bulk because 
$a (t, y)$ is roughly constant. Therefore, equation(\ref{zrs}) means that 
massless 
particles simply continue their path to the hidden brane and their faith 
depends on what happen to them there. At very high CM energy of UHECR 
interactions if charged particles can escape to the bulk photons are also 
dragged to the bulk and never come back. This is very similar to the 
conclusion of more precise and completely independent formulation of the 
previous section and shows that the logic behind them are consistent.

Next we consider the case of a general 2-brane model. Numerical solution of 
constrained 2-brane models in 
~\cite{kantres}~\cite{houribr} shows that for $\mu L \gtrsim 5$ the tension 
on both branes can be positive and very close to $\lambrs$. We use 
constraints on the Cosmological Constant and hierarchy to find ${\rho}'_0$ 
and ${\rho}'_L = {\rho}'_b$, the branes tension. We redefine them as 
${\rho}'_0 = 1 + \Delta 
{\rho}'_0$ and ${\rho}'_L = 1 + \Delta {\rho}'_L$. To solve hierarchy 
problem the following constraint must be fulfilled: 
\be
\frac {M^2_5}{M^2_{pl}} \sim N^2 \equiv \frac {n^2_L}{n^2_0} = 
\frac {{\rho}'_{{\Lambda}_0} (1-\cosh (\mu L)) + \sinh (\mu L)}{{\rho}'_{
{\Lambda}_L} (1-\cosh (\mu L)) + \sinh (\mu L)} \ll 1 \label {n2} 
\ee
This leads to:
\bea 
\Delta {\rho}'_0 & = & \frac {N^2 \biggl (1 - e^{-\mu L} + 
\Delta {\rho}'_L (1 - \cosh (\mu L) \biggr )}{1 - \cosh (\mu L)} - \nonumber \\
& & \frac {1 - e^{-\mu L}}{1 - \cosh (\mu L)} \label {deltrho0}
\eea
For a very small $N^2$ and $\Delta {\rho}'_L \lesssim 1$, the first term in 
(\ref{deltrho0}) is ${\mathcal O} (N^2)$ and:
\be 
\Delta {\rho}'_0 \approx - \frac {1 - e^{-\mu L}}{1 - \cosh (\mu L)} \approx 
- \frac {1}{1 - \cosh (\mu L)} \approx 2e^{-\mu L} \label {deltrho0p}
\ee
Using $\dot {a}^2_L / {a}^2_L = H^2$ where $H$ is the Hubble Constant on the 
visible brane~\cite{houribr}:
\bea
& & \hspace{-1cm}\Delta {\rho}'_L = \frac {1}{2 N^2 \sinh (\mu L)} \nonumber \\
& & \biggl [(1 - \cosh (\mu L)) \frac {2 H^2}{{\mu}^2} + e^{-\mu L} - 1 
\pm \nonumber \\
& & \biggl (\biggl((1 - \cosh (\mu L)) \frac {2 H^2}{{\mu}^2} + e^{-\mu L} - 
1 \biggr )^2 + \nonumber \\
& & N^2 \sinh (\mu L) (\frac {2 H^2}{{\mu}^2} + 2 e^{-\mu L})\biggr )^{\frac{1}{2}} \biggr ]\label {deltrhol}
\eea
In (\ref{deltrhol}) the solution with plus sign gives $\Delta {\rho}'_L 
\approx -2$ which deviates from our first assumption $|\Delta {\rho}'_L| < 1$ 
and leads to a negative tension on the visible brane like static RS model. 
The solution with negative sign is $\Delta {\rho}'_L \approx 2e^{-\mu L}$
and both branes have positive tension close to $\lambrs$.

\begin{figure}[t]
\begin{center}
\psfig{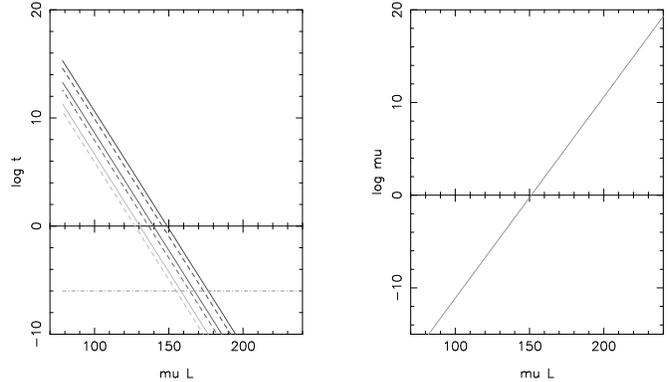}
\caption{Left: Propagation time for relativistic particles in 2-brane model. 
Description of the curves is the same as Fig.\ref{fig:rstime}. 
Right: Parameter $\mu (eV)$ as a function of $\mu L$. It is roughly 
independent of $M_5$.\label {fig:gtime}}
\end{center}
\end{figure}

Using these values for the tensions in (\ref{aa}) to (\ref{ndef}), we can 
find the time delay for this class of brane models:
\bea
\Delta t_{propag} & = & \frac {4}{\mu (8 |\Delta 
{\rho}'_0 + {\mathcal C}'|)^{\frac {1}{4}}} {\mathcal F} (\alpha, Q) 
\label {tprog2} \\
\alpha & = & 2 \arctan \sqrt \frac {q (z_- - z_0)}{p (z_0 - z_+)} \label 
{alphdef} \\
Q & = & \frac {1}{2} \biggl (2 + \frac {2}{pq} \nonumber \\
& &\hspace{-1cm}\biggl [\frac {{\mathcal C}'({\mathcal C}' - {\theta}^2) + 4 (2 + 
\Delta {\rho}'_0 + {\mathcal C}')(\Delta {\rho}'_0 + {\mathcal C}')}
{(\Delta {\rho}'_0 + {\mathcal C}')^2} \biggr ] \biggr)^{\frac{1}{2}} 
\nonumber \\
& & \label {qdef} \\
p^2 & \equiv & \biggl ( \frac {{\mathcal C}'}{\Delta {\rho}'_0 + 
{\mathcal C}'} - z_-\biggr )^2 + r^2 \\
q^2 & \equiv & \biggl ( \frac {{\mathcal C}'}{\Delta {\rho}'_0 + 
{\mathcal C}'} - z_+\biggr )^2 + r^2 \\
r^2 & \equiv & -\biggl ( \frac {{\mathcal C}'}{\Delta {\rho}'_0 + 
{\mathcal C}'} \biggr )^2 - \biggl (\frac {2 + \Delta {\rho}'_0 + 
{\mathcal C}'}{\Delta {\rho}'_0 + {\mathcal C}'} \biggr ) \label {pqr} \\
-{\mathcal C}' & = & \Delta {\rho}'_0 + \frac {2}{z_0} (N^2 + \Delta {\rho}'_0) 
\label {h2cond}
\eea
In (\ref{alphdef}), $z_0 = e^{\mu L}$ and, $z_{\pm}$ are the roots of velocity
amplitude.
The result of the numerical calculation of the time delay for these models 
is shown in Fig.\ref{fig:gtime}. We conclude that for these models the 
time delay can not constraint the value of $M_5$, but put constraint on 
$\mu L$, roughly speaking the warp factor. This means that by adjusting the 
warping, it is always possible to make the time delay too short to be 
observable at present air-shower detectors. Nevertheless, in these models 
$\mu$ is not arbitrary and depends on $\mu L$. On the other hand, in 
many radion models for stabilizing the bulk, $\mu$ is related to the mass of 
radion and its value must be in the range requested by the models for 
stabilization. Therefore comparing the results presented here permits to 
investigate which type of stabilization and particle physics behind them 
can satisfy all the constraints. Test of more models can be found in 
~\cite{houriescape}
\section{conclusion}
In this proceedings we tried to show that physics in future colliders and 
air-shower detectors is crowded by low energy effects and great efforts 
must be dedicated to extract information about new physics hidden deep 
inside collision events. 

We investigated the effect of the 
escape of partons to a warped bulk. We 
showed that it can significantly reduces the total cross-section and multiplicity - number of final state particles emitted into the 
brane. To arrive to this conclusion we needed to use an evolution equation 
to related the low energy observables to high energy parton (gluon) 
distribution in hadron-hadron collisions. We used the recently developed 
phenomenological model {\it Color Glass Condensate} and extended it to 4+1 
warped space-times with brane boundaries and static Randall-Sundrum metrics. 

Inspired by the observation that relativistic particles can be emitted to 
the bulk, we have also used a simple classical model of propagation of 
relativistic particles in bulk for more general type of brane models and 
obtained constraints both for the fundamental scale of gravity and for 
parameters of the brane models.
\Acknow {I would like to thank the organizers of the Cosmion-2004 
conference and very specially Dr. K. Belotsky and Pr. M. Khlopov.}

\end{document}